\pdfoutput=1
\documentclass[12pt]{article}
\usepackage{epsf,latexsym,rotate}
\usepackage{graphicx}
\usepackage[all]{xy}
\usepackage{amsfonts,amssymb} 
\epsfverbosetrue
\newcommand{\bb}{\begin{eqnarray}}
\newcommand{\ee}{\end{eqnarray}}
\newcommand{\eee}{\nonumber\end{eqnarray}}
\newcommand{\qq}{\quad}

\begin{document}

\font\twelve=cmbx10 at 13pt
\font\eightrm=cmr8

\thispagestyle{empty}

\begin{center}
${}$
\vspace{3cm}

{\Large\textbf{Strong lensing with positive cosmological constant}} \\

\vspace{2cm}

{\large
Thomas Sch\"ucker\footnote{also at Universit\'e de Provence,
thomas.schucker@gmail.com} (CPT\footnote{Centre de Physique
Th\'eorique\\\indent${}$\qq\qq CNRS--Luminy, Case
907\\\indent${}$\qq\qq F-13288 Marseille Cedex 9\\\indent${}$\qq
Unit\'e Mixte de Recherche (UMR 6207)
du CNRS et des Universit\'es Aix--Marseille 1 et 2\\
\indent${}$\qq et Sud
Toulon--Var, Laboratoire affili\'e \`a la FRUMAM (FR 2291)}) }

\vspace{3cm}

{\large\textbf{Abstract}}
\end{center}
Strong lensing by an isolated spherically symmetric mass distribution is considered in presence of a positive cosmological constant. Deflection angles and time delay are computed and compared to the multiple image of the quasar SDSS J1004+4112.
\vspace{2cm}

\vskip 1truecm

\noindent CPT-P034-2008\\
\vspace{2cm}
\vfil\eject

Last september Rindler \& Ishak \cite{ri} corrected the wide held belief that a cosmological constant does not change the deflection angle of light propagating near an isolated spherically symmetric mass distribution. This means that we must reconsider the theory of biangles. Biangles are one of the charming features of Riemannian geometry as they do not exist in Euclidean geometry. Like the theory of triangles, the theory of biangle relates angles and lengths, see figure 1. But relativity teaches us that distances have no physical meaning and are to be replaced by (proper) time of flight measurements of photons. 
\bigskip
\begin{center}
\begin{tabular}{c}
\xy
(0,0)*{}="L";
(60,0)*{}="S";
(-45,-13)*{}="T";
(-45,-13)*{\bullet};
(60,0)*{\bullet};
(0,10)*{r'};
(1,-16,5)*{r};
(0,10)*{\backslash};
(1,-16.5)*{\backslash};
(6,10)*{\tau'};
(7,-15.5)*{\tau};
"S"; "T" **\crv{(0,-20)};
"S"; "T" **\crv{(-10,20)};
(50,-3.3)*{}; (50,2.7)*{} **\crv{(47.5,0)};
(45,-0.5)*{\gamma  };
(-33,-3.6)*{}; (-31.5,-14.5)*{} **\crv{(-29.5,-8)};
(-26,-8)*{\beta };
(-17,-23)*{};
(75,0)*{};
\endxy
\end{tabular}
\linebreak\nopagebreak
{Figure 1: A biangle}
\end{center}
\bigskip
Indeed, almost 25 years ago,
on october 21st 1983 took place in S\`evres, on the western outskirts of Paris, the official funeral of the meter. The office was celebrated in a strict intimacy, but it is difficult to overestimate the loss caused to physics  by the disappearance of the meter. Announced in 1915 by Albert Einstein, experimentalists took several decades to authorise the funeral: ``The 17th Conf\'erence G\'en\'erale des Poids et Mesures decides: the metre is the length of the path travelled by light in vacuum during a time interval of 1/299 792 458 of a second.'' (http://www.bipm.org/en/CGPM/db/17/1/) Hence we measure lengths in nano-seconds, one nano-second being about one foot.

\begin{figure}[h]
\addtocounter{figure}{1}
\begin{center}
\includegraphics[viewport=0cm 0cm 19cm 15.5cm,width=11.5cm, height=9.5cm]{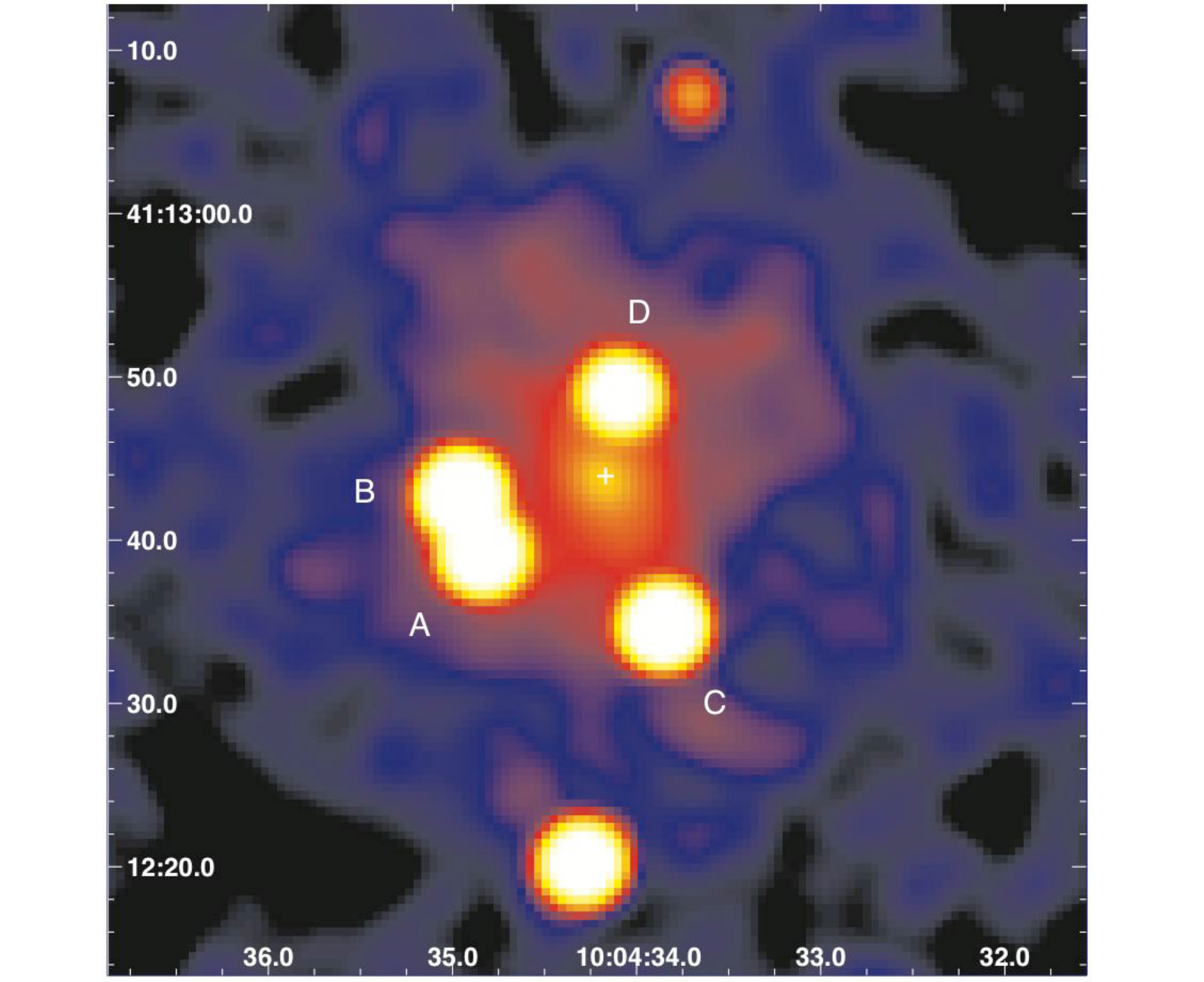}
\caption{The multiple imaged quasar SDSS J1004+4112 (courtesy of Ota et al. \cite{ot})}
\end{center}
\end{figure}

The first biangle was observed in 1968 by I. I. Shapiro. Its vertices are Earth and Mercury with a travel time $\tau$ of some 10 minutes and a time delay of 120 $\mu $s as one edge grazes the sun. Here we will be interested in biangles of cosmological scale, $\tau\sim 10^{10}$ years, figure 2. This picture, taken from Ota et al. \cite{ot}, shows four images  ($A$, $B$, $C$, $D$) of the same quasar lensed by a cluster of galaxies visible only via X-rays and whose center is indicated on the picture by a cross. For simplicity let us ignore the images $A$ and $B$ and let us pretend that the images $C$ and $D$ are aligned with the center of the cluster, which we then may treat as spherical. The biangle we are interested in, figure 3, lies in the plane orthogonal to the picure
and defined by the images $C$ and $D$. The position of terra nostra is indicated by $T$, of the cluster = lens by $L$ and of the quasar = source by $S$.  $\alpha $ and $\alpha '$ are the physical angles between the images and the lens. They are measured in 
nano-seconds over nano-seconds and appear directly on the picture. The coordinate distances $r_T$ and $r_S$ are radial coordinates of polar coordinates centered at the lens. The coordinate distances will be computed from the measured red-shifts of cluster and quasar,  $z_L$ and $z_S$. Let $M$ be the mass of the cluster.
\begin{center}
\begin{tabular}{c}
\xy
(0,0)*{}="L";
(60,0)*{}="S";
(-45,-13)*{}="T";
(-45,-13)*{\bullet};
(60,0)*{\bullet};
(62.5,3)*{S};
(50,5.5)*{D};
(50,-6)*{C};
(-47,-9.5)*{T};
(3.9,-2.8)*{L, M};
(6,10)*{\tau'};
(7,-15.5)*{\tau};
(0,0)*{\bullet};
"L"; "S" **\dir{-}; 
"S"; "T" **\crv{(0,-20)};
"S"; "T" **\crv{(-10,20)};
"L"; "T" **\dir{-}; 
(-33,-9.6)*{}; (-31.5,-14.5)*{} **\crv{(-30.5,-12)};
(-29,-8.4)*{}; (-31.5,-3.3)*{} **\crv{(-29,-5)};
(-25,-11)*{\alpha  };
(-24,-3)*{\alpha '};
(0,14)*{};
(25,-2.2)*{r_S};
(-15,-6.7)*{r_T};
(-17,-23)*{};
\endxy
\end{tabular}\\
{Figure 3: A double image}
\end{center}
 Ota et al. \cite{ot} observed the following values for angles, red-shifts and mass and last october Fohlmeister et al. \cite{fo} added a lower limit on the time delay, the jet lag of the photons:  
\bb \alpha =10''\,\pm\,10\% ,&
 z_L=0.68\ ,&M=5\cdot10^{13}M_\odot \,\pm\,20\%
 \nonumber\\
\alpha ' =\ 5''\,\pm\,10\% ,& z_S=1.734,&
\tau'-\tau>5.7\ {\rm y\ (oct.\ '07)}.
\eee
The proper time here is of course meant with respect to an observer on Earth. The jet lag being continuously monitored, this lower limit grows continuously and today it already exceeds six years.

Naturally we would like to confront observation and theory by taking into account Rindler \& Ishak's correction \cite{ri}. To this end we make the following simplifying assumptions:
\begin{itemize}
\item
Spatially flat $\Lambda CDM$ can be trusted to convert red-shifts into angular distances with respect to the Earth, which we denote by $d_L$ and $d_S$ respectively. We use $\Lambda < 1.77 \cdot 10^{-52}\ {\rm m}^{-2}$ in order to avoid negative dust densities.
\item 
We take the cluster to be static and spherically symmetric, see figure 4.
\item
We ignore all other masses in the universe.
\item
We ignore the velocities of the Earth and the quasar with respect to the cluster.
\end{itemize}
\begin{figure}[h]
\addtocounter{figure}{1}
\begin{center}
\vspace{-4cm}
\includegraphics[width=9.5cm, height=13.5cm]{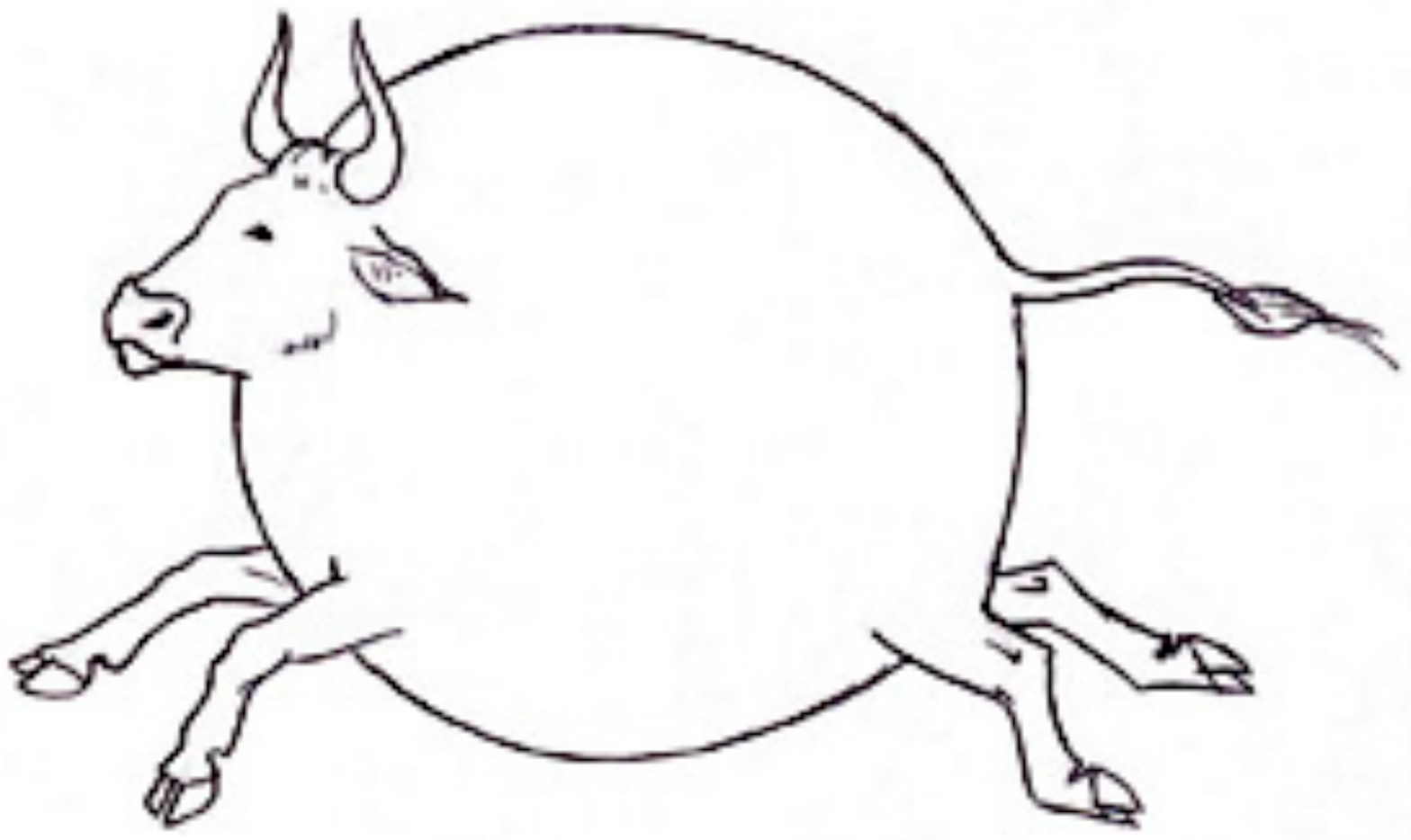}
\vspace{-4cm}
\caption{A spherical cow}
\end{center}
\end{figure}
In other words we allow ourselves to do the first part of the calculation with Friedmann's solution, the second part with Kottler's (or Schwarzschild - de Sitter's) solution, both however with the same cosmological constant $\Lambda $. Then we have \cite{ts}
\bb d_L=r_T,\qq d_S=\frac{r_T+r_S}{\sqrt{1-\Lambda r_S^2/3}},\qq
 \frac{r_T}{r_S} \,\sim\,\frac{4GM}{\alpha \alpha 'r_T}\,(1-\Lambda r_T^2/3)-1,\eee
 and \cite{sz}
\bb  \tau'-\tau&\sim&\sqrt{1-\Lambda r_T^2/3}\,GM
\left[ 
{\textstyle\frac{1}{2}\,}\frac{\alpha ^2-\alpha '^2}{1-\Lambda r_T^2/3}\,\frac{r_T}{GM} \,
\left( 1+\,\frac{r_T}{r_S} \right) 
+\,4\ln\frac{\alpha }{\alpha '}
 \right.\nonumber \\ &&\qq\qq
\left.-{\textstyle\frac{3}{2}}\,  
\left(1-\frac{\Lambda}{3}  r_T^2\right)\left( \frac{1}{\alpha '^2}-\frac{1}{\alpha ^2} \right)\frac{GM}{\sqrt{\Lambda /3}\,r_T^2}  \left( {\rm
arctanh} \sqrt{\frac{\Lambda}{3}}\,r_T +
{\rm
arctanh} \sqrt{\frac{\Lambda}{3}}\,r_S \right)
 \right] .\eee
The approximation indicated by $\sim$ is to leading order in three parameters: the Schwarzschild radius of the cluster divided by the peri-cluster, the angle $ \alpha $, and $\sqrt{\Lambda /3}$ times the peri-cluster. In our example, SDSS J1004+4112, all three parameters are of the order of $10^{-5}$ and the approximation is well justified. Notice that before Rindler \& Ishak's correction \cite{ri}, the above formula relating angles and lengths in the biangle was independent of the cosmological constant. 

Fitting the only parameter we have, i.e. the cosmological constant, to the observed values of the angles, red-shifts and cluster mass together with their error bars yields:
\bb 
\Lambda = 1.47^{+0.3}_{-0.9}
 \cdot 10^{-52}\ {\rm m}^{-2},\qq \tau'-\tau= 19.6 ^{+7.6}_{-7.3}
\  {\rm y}.
  \eee
  For details see table 1.
The upper limits correspond to vanishing dust density in Friedmann's solution. Note that the fit  is compatible with the present experimental value, $\Lambda =(1.36\pm 0.3)\cdot10^{-52}\ {\rm m}^{-2}$ and with the present lower limit on the time delay.

\begin{table}
\begin{center}  
\begin{tabular}{|c||c|c|c|c|c|c|c|c|c|c|c|}
\hline
$M\pm 20\%$&$\pm 0$&$ +$&$-$&$-$&$-$&$-$&+&+&$\!\!+8.5\%\!\!$&+&+ \\ 
\hline
$\alpha \pm 10\%$&$\pm 0$& +&+&$-$&$-$&+&+&$-$&$-$&$\!\!-0.6\%\!\!$&$-$
                   \\ 
                   \hline
$\alpha' \pm 10\%$&$\pm 0$& +&+&+&$-$&$-$&$-$&+&$-$&$-$&$\!\!-0.6\%\!\!$
                   \\ 
\hline\hline
$r_T\, [10^{25}\,{\rm m}]$&7.9 & 7.9&{\bf 6.5}&7.2&7.9&7.2&8.5& 8.5&8.9&{\bf 8.9}&8.9  \\ 
\hline
$r_S\, [10^{25}\,{\rm m}]$&6.6 & 6.6&{\bf 4.9}&5.7&6.6&5.7&7.7&7.7&8.4&{\bf 8.4}&8.4  \\ 
\hline
$\!\!\Lambda [10^{-52}\,{\rm m}^{-2}]\!\!$&1.5 & 1.5&{\bf 0.6}&1.1&1.5&1.1&$1.7$&$1.7$&${\bf 1.77}$&{\bf 1.77}&{\bf 1.77}    \\ 
\hline
$\!\!\tau'-\tau\, [{\rm years}]\!\!$ &19.6& 23.6& 18.9&$\!{\bf 12.3}\!$&15.8&22.6&$\!{\bf 27.2}\!$& 14.8&17.6&22.3&16.6  \\
\hline
\end{tabular} 
\end{center}
\caption{ The fit, extrema are bold face.}
\end{table} 

Our result for the time delay should be compared to Kawano \& Oguri's result \cite{ko},
$\tau'-\tau\,<\, 10$ years. They start from a non-spherical lens, thus including also the images $A$ and $B$, but assume a vanishing cosmological constant.

In any case, a time delay of 20 years is bad news for the old physicists among us: I seriously doubt that in 15 years I will still care about the jet lag of pre-historic photons.  

Before we can trust strong lensing as a new probe of the cosmological constant we must 
\begin{itemize}\item
remove the simplifying assumptions introduced above,
\item
 analyze more lensing systems.
 \end{itemize}
  The use of the Einstein-Straus solution \cite{es,sch} has been advocated by Ishak et al. \cite{ir}, version 2. This solution matches up Schwarzschild and Friedmann's  solutions and thereby avoids the use of two different solutions in the same calculation.  It takes proper account of the other masses in the universe and of the velocities of  Earth and source. The Einstein-Straus solution therefore circumvents our simplifying assumptions, apart from that of the spherical cow.  In a preliminary analysis Ishak et al. \cite{ir} find that the Einstein-Straus solution reduces the sensitivity of the deflection angles as a function of the cosmological constant by two orders of magnitude. Qualitatively this reduction is easy to understand: A positive cosmological constant adds a repulsive force, which grows with distance, and one cannot invoke asymptotic flatness anymore to justify the neglecting of the other masses in the universe. If we take them into account and if we distribute them uniformly in space, they will screen the long-range repulsion of our central cluster.

Those of you believing in dark energy might want to know how it modifies the Schwarzschild solution and deflection angles of photons. Be aware that there is no answer. Finelli et al. \cite{fi} present a partial parameterization of our ignorance. Even, more explicite models of dark energy, like quintessence, do not allow us to predict the deflection angle \cite{cw}.
\\[0.2cm]
{\it Acknowledgement:} I thank Naomi Ota for his kind permission to reproduce figure 2.

\end{document}